\documentclass[conference]{IEEEtran}
\DeclareUnicodeCharacter{2032}{\ensuremath{'}}

\IEEEoverridecommandlockouts
\usepackage{cite}
\usepackage{amsmath,amssymb,amsfonts}
\usepackage{graphicx}
\usepackage{textcomp}
\usepackage{xcolor}
\usepackage{algorithm}
\usepackage{algpseudocode}
\usepackage{array}
\usepackage{hyperref}
\usepackage{moresize}
\usepackage{makecell}
\usepackage{booktabs}

\newcommand{\tpad}{\rule{0pt}{3.5ex}}       
\newcommand{\tpadtight}{\rule{0pt}{0.0ex}}  

\def\BibTeX{{\rm B\kern-.05em{\sc i\kern-.025em b}\kern-.08em
    T\kern-.1667em\lower.7ex\hbox{E}\kern-.125emX}}
\begin{document}

\title{Co-optimization of codon usage and mRNA secondary structure using quantum computing}

\author{\IEEEauthorblockN{Dimitris Alevras}
\IEEEauthorblockA{\textit{IBM Quantum}\\
New York, USA \\
alevras@us.ibm.com}
\and
\IEEEauthorblockN{Mihir Metkar}
\IEEEauthorblockA{\textit{Moderna}\\
Cambridge, USA \\
mihir.metkar@modernatx.com}
\and
\IEEEauthorblockN{Triet Friedhoff}
\IEEEauthorblockA{\textit{IBM Quantum}\\
New York, USA \\
triet.nguyen-beck@ibm.com}
\and
\IEEEauthorblockN{Jae-Eun Park}
\IEEEauthorblockA{\textit{IBM Quantum}\\
New York, USA \\
parkje@us.ibm.com}
\and
\IEEEauthorblockN{Mariana LaDue}
\IEEEauthorblockA{\textit{IBM Quantum}\\
New York, USA \\
mariana.ladue@ibm.com}
\and
\IEEEauthorblockN{\qquad\qquad\qquad}
\and
\IEEEauthorblockN{Vaibhaw Kumar}
\IEEEauthorblockA{\textit{IBM Quantum}\\
New York, USA \\
vaibhaw.kumar@ibm.com}
\and
\and
\IEEEauthorblockN{Wade Davis}
\IEEEauthorblockA{\textit{Moderna}\\
Cambridge, USA \\
wade.davis@modernatx.com}
\and
\IEEEauthorblockN{Alexey Galda}
\IEEEauthorblockA{\textit{Moderna}\\
Cambridge, USA \\
alexey.galda@modernatx.com}
\and
\IEEEauthorblockN{\qquad\qquad\qquad}
}

\maketitle

\begin{abstract}

Co-optimizing mRNA sequences for both codon optimality and secondary structure is crucial for producing stable and efficacious mRNA therapeutics. Codon optimization, which adjusts nucleotide sequences to enhance translational efficiency, inherently influences mRNA secondary structure -- a key determinant of molecular stability both in-vial and in-cell. Because both properties are governed by the same underlying sequence, optimizing one directly impacts the other. To address this interdependence, we introduce a novel variational framework that simultaneously optimizes codon usage and secondary structure. Our method employs a dual-objective function that balances the codon adaptation index (CAI) and minimum free energy (MFE), incorporating variational parameters for codon selection. Leveraging a hybrid quantum-classical computational strategy and building on prior advancements in quantum algorithms for secondary structure prediction, we effectively navigate this complex optimization space. We demonstrate the feasibility of executing this end-to-end workflow on real quantum hardware, using IBM’s 127-qubit Eagle processor. We validate our approach through both simulations and quantum hardware experiments on sequences of up to 30 nucleotides. These results highlight the potential of our framework to accelerate the design of optimal mRNA constructs for therapeutic and research applications.

\end{abstract}

\section{Introduction}
The rapidly advancing field of mRNA therapeutics, exemplified by the swift development of COVID-19 vaccines, highlights its potential to transform medical treatment modalities. This potential hinges on the ability to design therapeutic mRNAs that achieve high stability and translation efficiency. Balancing these elements is challenging due to the complex interactions between nucleotide sequences, base modifications, and RNA structures -- all of which together can affect mRNA degradation and translation kinetics~\cite{metkar2024tailor, beaudoin20105}.

Co-optimizing codon usage and secondary structure is an established strategy for improving mRNA performance, directly enhancing translational efficiency and thereby increasing protein expression\cite{mauger2019mrna, metkar2024tailor}. In addition, optimizing mRNA structure not only improves intracellular performance but also contributes to improved stability during storage and transport. However, due to the astronomical number of possible nucleotide sequences and secondary structures for any given protein, designing therapeutic mRNAs requires specialized algorithms capable of efficiently navigating this complexity~\cite{andronescu2008rna}. State-of-the-art machine learning mRNA design algorithms integrate advanced biochemical principles with cutting-edge computational methods to address this multi-objective optimization
problem~\cite{leppek2022combinatorial, li2024codonbert, vostrosablin2024mrnaid, wayment2021theoretical, zhang2023algorithm, gu2024derna}. A comprehensive review of these algorithms can be found in Ref.~\cite{ward2025mrna}. Recent work is also beginning to explore quantum computing methods for mRNA design, including early demonstrations of codon optimization frameworks in hybrid quantum-classical settings~\cite{zhang2024resource, chung2024quantum}. While prior efforts have focused on simulated quantum workflows, we extend this work further by demonstrating real-hardware execution of the quantum subroutine, validating its feasibility on today’s superconducting quantum processors.

In this work, we present a hybrid quantum-classical approach for solving the multi-objective optimization problem of mRNA design. Beyond simulation, we demonstrate that our method can be successfully executed on current superconducting quantum hardware.

\section{Methods}

This Section presents a variational algorithm that iteratively solves a parameterized codon optimization problem and uses its solution as input to the secondary structure prediction problem. The codon optimization output is used to compute the {\em codon adaptation index} (CAI)~\cite{sharp1987codon}, while the solution to the secondary structure prediction problem is used to evaluate the {\em minimum free energy} (MFE)~\cite{tinoco1971estimation} of the structure. The algorithm optimizes a composite objective function defined as a weighted sum of CAI and MFE:
\begin{equation}\label{eq:f}
    f(\mathbf{\theta}) = \alpha \ \text{CAI}(\mathbf{\theta}) + \text{MFE}(\mathbf{\theta}) \ ,
\end{equation}
where $\alpha$ is a tunable parameter balancing the two terms.

The codon optimization problem considers a sequence of $n$ amino acids $A=\{a_1,a_2,\dots,a_n\}$, where each amino acid $a_i$ can be encoded by a set $C(a_i)$ of one to six synonymous three-nucleotide codons. The goal is to select one codon per amino acid such that the resulting nucleotide sequence optimizes a set of biological and structural criteria:
\begin{itemize}
\item GC content: the number of ``G" and ``C" nucleotides in the sequence.
\item Rarity score: a penalty based on how infrequently codons are used across known sequences.
\item Repeat score: a penalty for repeating nucleotides in adjacent codons.
\end{itemize}

The codon optimization is formulated as a binary linear program. Let $x_{i,j}$ be a binary variable that equals $1$ if codon $j$ is selected for amino acid $a_i$, and 0 otherwise. Each amino acid must be assigned exactly one codon, resulting in the constraint:
\begin{equation}
\sum_{j \in C(a_i) } x_{i,j} = 1 \quad \quad \forall~ i=1,\ldots ,n.
\end{equation}
The three objective terms are defined as follows (adapted from Ref.~\cite{fox_codon}):

\noindent {\bf GC content}\\
Let $g(j)$ and $c(j)$ be the number of ``G" and ``C'' bases in codon $j$, respectively. Then the GC content of codon $j$ is $gc(j) = g(j)+c(j)$, and the total GC content of the sequence is 
\begin{equation}
\sum_{i =1}^{n} \sum_{j \in C(a_i)} gc(j) x_{i,j}.
\end{equation}

\noindent {\bf Rarity score}\\
The rarity score seeks to penalize use of rare codons~\cite{plotkin2011synonymous}. The codon usage frequency varies by host system, and following Refs.~\cite{sharp1987codon, fox_codon} we use the logarithm function to penalize usage of infrequent codons. The frequencies used are for {\it H. sapiens}, as found in the {\tt python\_codon\_tables} library~\cite{codtab}. Let $p_j$ be the logarithm of the frequency of codon $j$. The total penalty contribution in the objective function is
\begin{equation}
\sum_{i=1}^n \sum_{j \in C(a_i)} p_j x_{i,j}.
\end{equation}
 
\noindent {\bf Repeat score}\\
Let $r(j,k)$ denote the number of repeated nucleotides for a sequence consisting of codons $j$ and $k$. Following Ref.~\cite{fox_codon}, we define this function as the square of the number of repeated nucleotides minus 1. If codon $j$ encodes amino acid $a_i$ at position $i$, and codon $k$ encodes the same amino acid at position $i+1$, then a penalty term $r(j,k) x_{i,j} x_{i+1,k}$ is added to the objective function. The total penalty is given by:
\begin{equation}
\sum_{i=1}^{n-1} \sum_{j \in C(a_i)} \sum_{k \in C(a_{i+1})} r(j,k) x_{i,j} x_{i+1,k}.
\end{equation}

Each term in the objective function is assigned a weight, which serves as a variational parameter for the optimization algorithm. 
The codon optimization problem can then be formulated as follows:
\begin{equation}
\begin{aligned}
    \min \ &
    \theta_c \sum_{i =1}^{n} \sum_{j \in C(a_i)} gc(j) x_{i,j} + \\
 &\theta_p \sum_{i=1}^n \sum_{j \in C(a_i)} p_j x_{i,j} + \\
 &\theta_r\sum_{i=1}^{n-1} \sum_{j \in C(a_i)} \sum_{k \in C(a_{i+1})} r(j,k) x_{i,j} x_{i+1,k} \\
 \text{subject to}& \\
 &\sum_{j \in C(a_i) } x_{i,j} = 1 \quad \quad \forall i \in \{1,\ldots ,n\}\,, \\
 &x_{i,j} \in \{0,1\} \ i=1,\ldots,n, \ \ j \in C(a_i)\,,
 \end{aligned}
\end{equation}
where $\mathbf{\theta}=(\theta_c, \theta_p, \theta_r)$ is the initial vector of variational parameters. The output of the codon optimization problem is a nucleotide sequence defined as ${\{j \ \text{for } i=1,\ldots,n,\ \text{if } x_{i,j} =1\}}$, which serves as the input to the secondary structure prediction problem.

Following Ref.~\cite{alevras2024mrna}, we model mRNA secondary structure prediction as a binary optimization problem using mathematical programming. Each position $i \in \{1,\ldots,n\}$ in the mRNA sequence corresponds to one of the nucleotides (bases) from the alphabet $\{U,A,C,G\}$. A base pair between positions $i$ and $j$ is allowed if the nucleotides at those positions form one of the valid Watson-Crick or wobble pairs: $\{(AU), (UA), (CG), (GC), (GU),(UG)\}$. Following the formulation in Ref.~\cite{gusfield2019integer}, we define binary variables over quartets, which represent two consecutive base pairs (also known as stacked pairs). For indices $i$ and $j$, the quartet variable ${x(i,j,i+1,j-1)}$ is defined as:
$$x(i,j,i+1,j-1)=\left \{ 
\begin{array}{rl}
\!\!\!1\!\!\! & \text{if $(i,j)$ and $(i+1,j-1)$ are formed}\\
\!\!\!0\!\!\! & \text{otherwise.}
\end{array} 
\right .
$$ 

\begin{figure*}[t]
\centering
        \frame{\includegraphics[width = \textwidth]{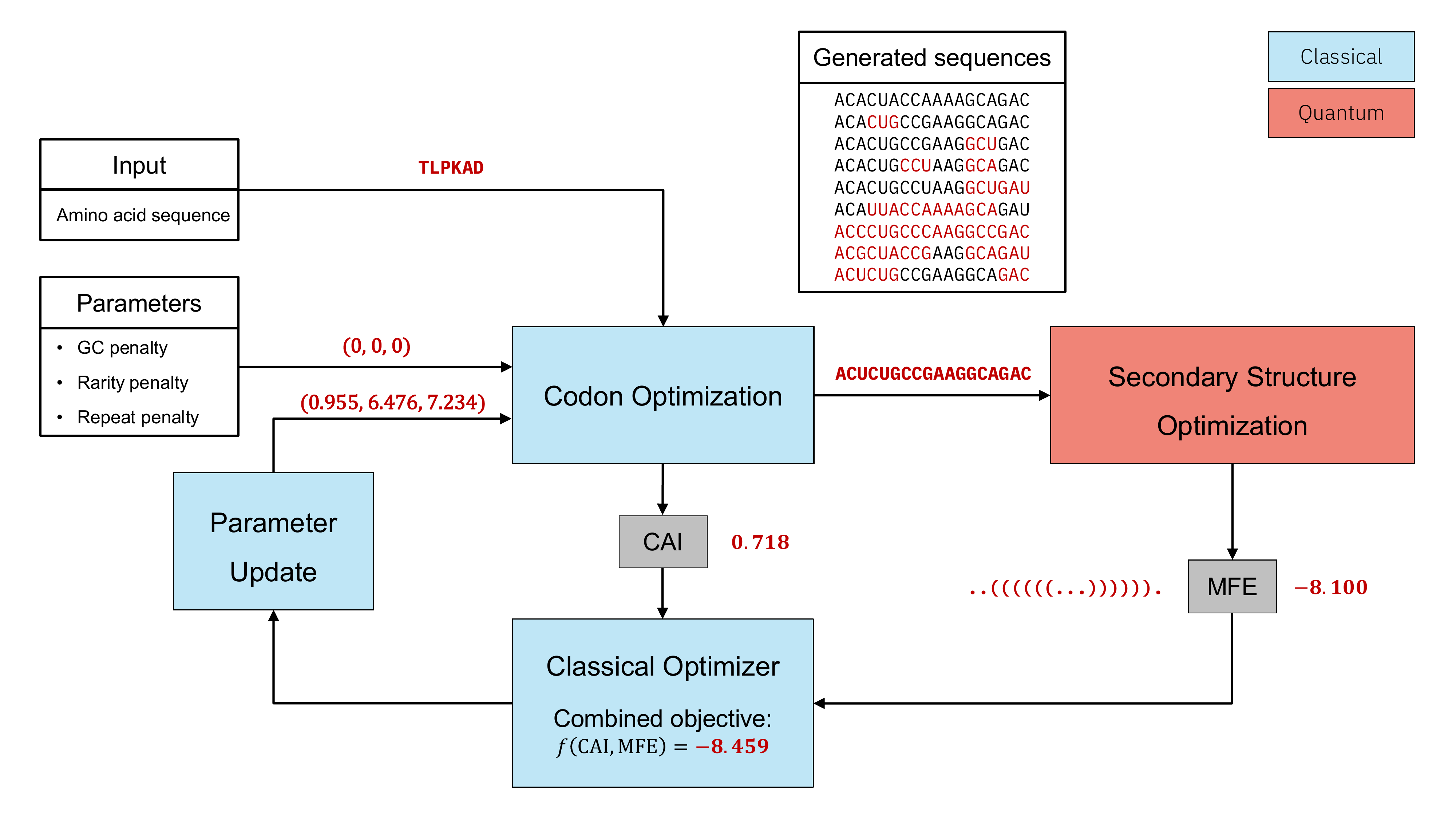}}
        \caption{Workflow of the hybrid mRNA optimization algorithm. Starting from an amino acid sequence, a classical optimizer iteratively updates codon selection parameters to maximize a combined CAI–MFE objective. Generated nucleotide sequences are evaluated using classical codon metrics and quantum-assisted secondary structure prediction.}\label{fig:flow}
\end{figure*}

These variables are generated only when the corresponding base pairs form valid quartets. To prepare the input data for the optimization formulation, the following preprocessing steps are applied:
\begin{itemize}
    \item Verify whether two base pairs can form a valid quartet (i.e., whether they can nest without violating pairing rules).
    \item Check whether any two base pairs share a common nucleotide.
    \item Detect crossing base pairs that would violate the planar structure of mRNA.
    \item Construct the list of all valid quartets, which define the binary variables of the problem.
    \item Identify and mark quartets that are mutually incompatible due to crossing.
\end{itemize}

Two quartets cannot be simultaneously selected if any of their constituent base pairs are crossing. Additionally, each nucleotide can participate in at most one base pair. These structural requirements give rise to constraints of the form:
$$x(i,j,i+1,j-1) + x(k,\ell,k+1,\ell-1) \le 1\,,$$
which ensure that conflicting quartets are not both active. Since the variables are binary, this linear constraint is equivalent to the quadratic condition:
$$x(i,j,i+1,j-1)  x(k,\ell,k+1,\ell-1) = 0\,,$$
which can alternatively be incorporated into the objective function as a penalty term. This penalizes the simultaneous selection of incompatible quartets, discouraging violations of structural feasibility. The choice between enforcing these relationships as hard constraints or incorporating them as soft penalties in the objective depends on the optimization method used. In our experiments, we employ {\tt CPLEX}~\cite{ibmlicense}, which exhibits better performance when explicit constraints are included in the problem formulation.

The objective function for mRNA secondary structure prediction incorporates several biologically motivated components:
\begin{itemize}
    \item {\bf Free energy of quartets:} Each variable, representing a valid quartet (stacked base pairs), is associated with a free energy contribution as defined in Ref.~\cite{turner-mathews}. These values are used to favor energetically favorable structures.
    \item {\bf Stability of consecutive quartets:} Consecutive quartets contribute to the overall stability of the secondary structure. To promote this, we include a reward term in the objective function proportional to the product of variables corresponding to adjacent quartets.
    \item {\bf Penalties for undesired motifs:} Certain structural features are biologically less favorable. In this work, we penalize structures ending in a $(UA)$ pair, following the terminal penalty described in Ref.~\cite{turner-mathews}. This is implemented by adding a penalty term for variables representing such configurations.
\end{itemize}

For notational convenience, we denote the $i^{th}$ quartet variable by $q_i$. Let $Q$ be the set of all valid quartets, $QC$ the set of all mutually crossing quartet pairs, $QS(q_i)$ the set of quartets that can be stacked with $q_i$, and $\mathit{QUA}$ the set of stacked quartets that terminate with a $(UA)$ end pair. These sets -- $Q$, $QC$, $QS(q_i)$, and $\mathit{QUA}$ -- are derived from the input sequence during preprocessing. Each quartet $q_i \in Q$ is formed by two consecutive base pairs $p^i_1$ and $p^i_2$. The MFE associated with $q_i$, denoted $e_{q_i}$, corresponds to the energy contributions of the stacked pair $p^i_1$ followed by $p^i_2$, as described in Ref.~\cite{turner-mathews}. If two quartets $q_i, q_j \in Q$ form a valid stack, their joint selection is rewarded by an energy term $r$. Conversely, if a stacking interaction ends with a $(U,A)$ pair, a penalty $p$ is incurred. These contributions are combined to define the overall objective function:
\begin{equation} \label{eq:obj}
    \min \sum_{q_i \in Q} e_{q_i} q_i  +
r\sum_{q_i \in Q}{\sum_{q_j \in QS(q_i)}}q_i q_j +
p \sum_{q_i \in Q} \sum_{q_j \in \mathit{QUA}} q_i (1-q_j)\,.
\end{equation}

The constraints of the problem enforce that any two quartets $q_i, q_j \in Q$ that involve crossing base pairs cannot be simultaneously selected. This requirement is captured by the following inequality:
\begin{equation}\label{eq:constraint}
    q_i + q_j \le 1\,, \quad \quad \forall q_i, q_j \in QC\,.
\end{equation}
Together with the binary nature of the decision variables, this defines a quadratic binary optimization problem. 

With both the codon optimization and secondary structure prediction problems formally defined, we now turn to the algorithmic strategy for solving the combined problem.

\begin{table*}[htbp]
\centering
\caption{Simulation results for various amino acid sequences.}
\scriptsize
\begin{tabular}{@{}lc p{4.0cm}c l c c c@{}}
\toprule
\textbf{AA Sequence} & \textbf{AA} & \textbf{Nucleotide Sequence} & \textbf{Nuc. Seq.} & \textbf{Folding} & \textbf{Qubits} & \textbf{Iterations/} & \textbf{Success} \\
\textbf{} & \textbf{Length} &  & \textbf{Length} &  & \textbf{(min -- max)} & \textbf{Fun. Evals} & \textbf{Rate} \\
\midrule
\tpad TLPKAD & 6 & \makecell[l]{ACUCUGCCUAAGGCGGAC} & 18 & ..((((((...)))))). & 2 -- 7 & 24/99 & 100 \\
\tpad IMQWIGCY & 8 & \makecell[l]{AUUAUGCAGUGGAUCGGCUGCUAC} & 24 & .....(((((......)))))... & 16 -- 22 & 13/53 & 100 \\
\tpadtight DRNKFHLRWD & 10 & \makecell[l]{GAUAGGAAUAAGUUCCACCUGAGG\\UGGGAC} & 30 & ...........((((((((...)))))))) & 27 -- 33 & 13/53 & 100 \\
\tpad YDDCAVNWCWVEY & 13 & \makecell[l]{UACGACGACUGCGCUGUGAACUGG\\UGCUGGGUCGAGUAC} & 39 & (((..(((((((((((.....))))))..))))).))). & 63 -- 89 & 13/53 & 80 \\
\tpad SKDPVRCVSIAEV & 13 & \makecell[l]{AGUAAGGAUCCAGUGAGAUGCGUG\\AGCAUCGCUGAGGUC} & 39 & ......((((((((((..(((....))))))))).)))) & 41 -- 52 & 21/66 & 80 \\
\tpad UWKLCRIWYYRCDC & 14 & \makecell[l]{ACCUGGAAGCUGUGUCGCAUCUGG\\UACUACAGAUGCGACUGC} & 42 & ...........((((((((((((......)))))))))).)) & 67 -- 112 & 12/48 & 80 \\
\tpad MLCDKVLKMHWMFHRE & 16 & \makecell[l]{AUGCUGUGCGACAAGGUGCUGAAG\\AUGCACUGGAUGUUCCACAGAGAG} & 48 & ...(((((.((((.(((((.......)))))...)))).))))).... & 75 -- 104 & 12/48 & 70 \\
\bottomrule
\end{tabular}
\label{tab:simres}
\end{table*}

To optimize the composite objective function $f(\mathbf{\theta})$, Eq.~\ref{eq:f}, we employ a classical optimizer - specifically, the Nelder–Mead algorithm~\cite{nelder1965simplex}. The optimization requires three inputs: an amino acid sequence $A$, a tunable weight parameter $\alpha$ (set to $-0.5$ in our optimization runs), and an initial guess for the parameter vector $\mathbf{\theta}$.

Each evaluation of $\mathbf{\theta}$ involves the following sequence of computations. First, the codon optimization problem is solved with the current parameters, yielding a candidate nucleotide sequence. We then compute its CAI and use it as input to the secondary structure prediction module. The predicted structure is scored via MFE, and the resulting value is used to evaluate the composite objective $f(\mathbf{\theta})$.

The codon optimization problem is solved exactly using the {\tt CPLEX} solver~\cite{ibmlicense}, as it is not NP-hard and admits efficient classical solutions. In contrast, secondary structure prediction is more computationally intensive and is handled by a CVaR variational quantum algorithm, following Ref.~\cite{alevras2024mrna}. The CVaR objective improves robustness by targeting the low-energy tail of the distribution, filtering out noise-induced outliers—an approach particularly beneficial on near-term quantum processors with broad sampling distributions. Prior applications in quantum chemistry and variational eigensolvers have shown that CVaR enhances convergence speed and stability~\cite{barkoutsos2020improving}. To reduce redundant quantum evaluations across iterations, we cache previously encountered nucleotide sequences and their corresponding optimal foldings.

\floatname{algorithm}{Function}
\begin{algorithm}
    \caption{Composite objective function}\label{cod:func}
    \begin{algorithmic}
        \Function{F}{$A, \theta$}
          \State $s \gets \text{CO}(A, \theta)$ \Comment{Codon Optimization}
          \State $f_{\text{CAI}} \gets \text{CAI}(A, s)$
          \State $dot\_b \gets \text{SSP}(s)$ \Comment{Secondary Structure Prediction}
          \State $f_{\text{MFE}} \gets \text{MFE}(s, dot\_b)$
          \State \Return $f \gets \alpha \cdot f_{\text{CAI}} + f_{\text{MFE}}$
        \EndFunction
    \end{algorithmic}
\end{algorithm}

The CAI for a given nucleotide sequence 
$\mathcal{C} = \{c_1, \dots, c_n \}$, which encodes the amino acid sequence $A=\{a_1,\ldots,a_n\}$, is computed as follows. The codon usage table provides the frequency with which each codon is used to express a given amino acid. These frequencies are normalized to codon-specific scores $w_i$, where the most frequently used codon is assigned a score of 1, and all other synonymous codons receive a score between 0 and 1. The CAI, introduced in Ref.~\cite{sharp1987codon}, is then defined as the geometric mean of the scores across all codons in the sequence:
$$CAI = \left ( \prod_{c_i \in \cal{C}} w_i \right )^{\frac{1}{n}}$$

The MFE of a given mRNA secondary structure is computed using the {\tt eval\_structure} method of {\tt fold\_compound} class from the {\tt viennaRNA} Python package~\cite{vienna}.

\section{Results}

\subsection{Simulation Results}
We evaluate the performance of our hybrid optimization framework through both classical simulation and real quantum hardware experiments. We begin by demonstrating its behavior on short amino acid sequences using classical simulation. Fig.~\ref{fig:flow} illustrates the workflow on a 6-amino-acid example.

Starting from the amino acid sequence {\tt TLPKAD} and initial parameters $\mathbf{\theta} = (0,0,0)$, the algorithm generates candidate nucleotide sequences through iterative optimization. The final output sequence, \texttt{ACUCUGCCGAAGGCAGAC}, is obtained via {\tt CPLEX} with optimized parameters $\mathbf{\theta} = (0.955,6.476,7.234)$. The resulting sequence achieves a codon adaptation index of ${\mathrm{CAI}=0.718}$ and serves as the input to the quantum secondary structure prediction algorithm, which produces the mRNA folding ``{\tt..((((((...)))))).}'' in the dot-bracket notation. The MFE associated with this structure is ${\mathrm{MFE}=-8.100}$, and the corresponding composite objective value is ${f(\mathrm{CAI},\mathrm{MFE})=-8.459}$. Notably, the predicted folding matches the classically computed optimal structure. The full optimization run, using the Nelder-Mead algorithm, converged in 27 iterations and required 106 function evaluations. 

Table~\ref{tab:simres} includes the input amino acid sequence and its length, along with the final nucleotide sequence and the predicted RNA secondary structure (in dot-bracket notation). For each instance, we report the minimum and maximum number of qubits required to solve the secondary structure prediction problem, the number of optimizer iterations and function evaluations, and the success rate across ten independent runs. The qubit range reflects the fact that the size of the structure prediction problem varies depending on the nucleotide sequence generated during optimization.

\subsection{Quantum Hardware Results}

We next validate the workflow on actual quantum hardware using IBM’s 127-qubit superconducting processors. Specifically, we executed the optimization algorithm on the \texttt{TLPKAD} sequence using IBM’s 127-qubit Eagle processor \texttt{ibm\_brussels}. Error suppression was applied via dynamical decoupling using the default 'XX' sequence, and the transpiled quantum circuit was executed with ${N_{\text{shots}}=2^{13}}$ bitstrings collected via Qiskit’s sampling primitive.

Execution on quantum hardware successfully reproduced the trajectory observed in simulation. Starting from the amino acid sequence, the algorithm identified an optimal nucleotide sequence and folded it into the classically predicted secondary structure during the early stages of optimization.

\begin{figure}
\centering
        \includegraphics[width = 0.53 \textwidth ]{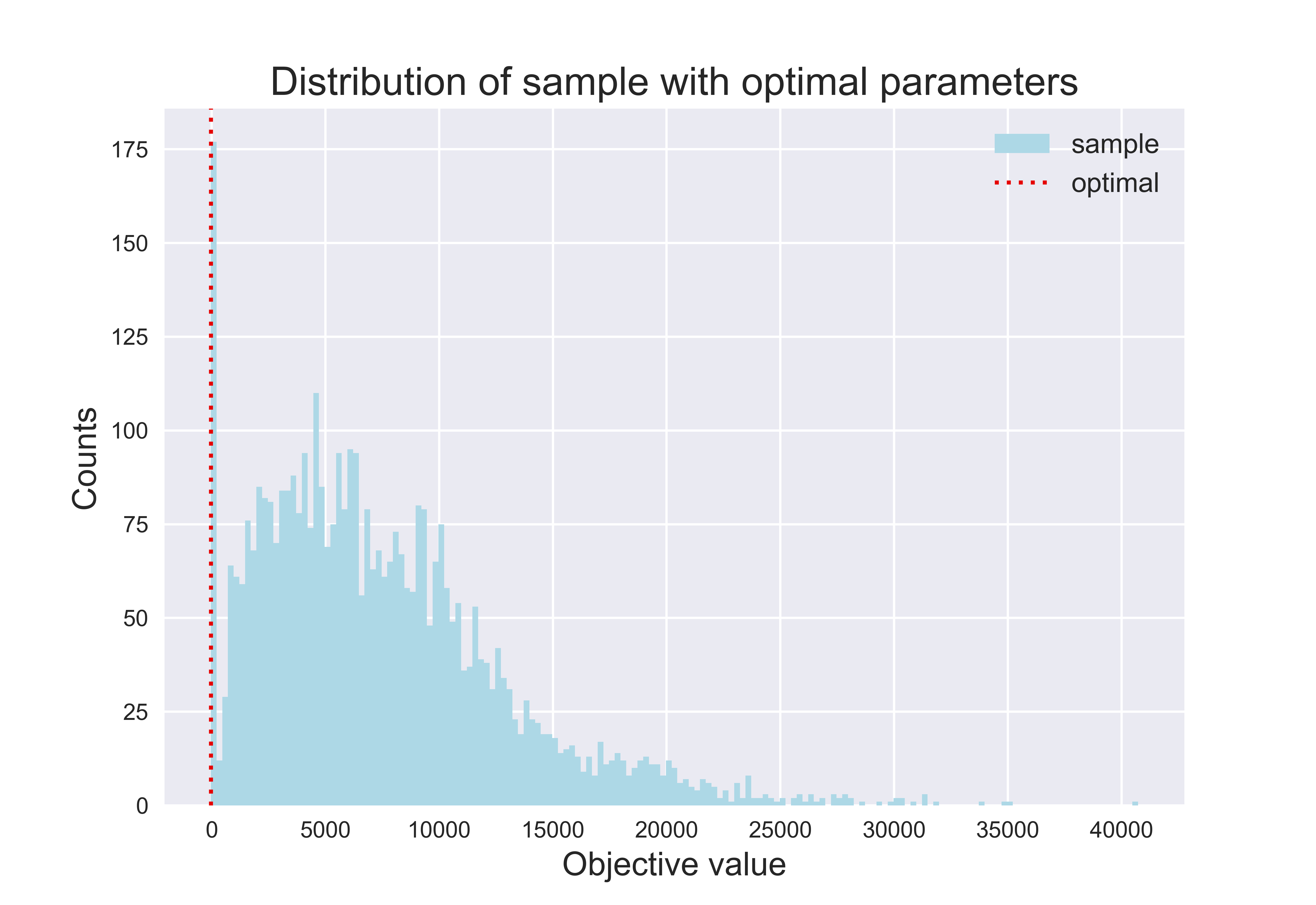}
        \caption{Sample distribution of objective values from \texttt{ibm\_fez} run with simulator-obtained optimal parameters.}\label{fig:hwsam}
\end{figure}

To further examine scalability and performance, we selected a longer 13-amino-acid sequence and ran the algorithm on the simulator to obtain optimal parameters. These were then used for a hardware sampling run on \texttt{ibm\_fez}, targeting the amino-acid sequence \texttt{YDDCAVNWCWVEY}. The resulting nucleotide sequence, \texttt{UACGACGACUGCGCUGUGAACUGGUGCUGGGUCGAGUAC}, was submitted for sampling, and the observed distribution of objective values, see Fig.~\ref{fig:hwsam}, demonstrates that the optimal structure was sampled with the highest frequency.

Together, these results demonstrate the feasibility of executing our hybrid algorithm on real quantum hardware. The high correspondence between simulation and hardware runs confirms the reliability of the quantum-assisted secondary structure prediction step. These findings establish an important milestone toward practical, biologically relevant quantum computation for mRNA design.

\section{Conclusions}

We presented a hybrid variational algorithm that jointly addresses codon optimization and mRNA secondary structure prediction. Our approach leveraged classical solvers for codon selection and coordination, while employing a quantum algorithm—simulated using classical backends—for the more complex secondary structure prediction task.

Initial experiments showed that the algorithm reliably identified biologically plausible solutions for short amino acid sequences. These results validated the feasibility of integrating quantum and classical methods within a unified optimization framework for mRNA design. Importantly, we demonstrated that the full workflow can be executed on real quantum hardware using IBM’s 127-qubit Eagle processor. The quantum algorithm consistently reproduced the results obtained via classical simulation, confirming correctness and representing a key milestone in biologically relevant quantum computation.

Future work will focus on scaling our approach to longer and biologically relevant mRNA sequences. As the number of codon choices and structural constraints increases, the optimization runtime grows accordingly, and quantum circuit depth may become a limiting factor on near-term hardware. We also aim to improve the efficiency of the classical optimizer, with particular emphasis on reducing the number of iterations and function evaluations required for reliable convergence.

To address these challenges, we are investigating several strategies, including decomposition of the secondary structure problem into modular subregions, integration of surrogate models to lower the evaluation cost during optimization, and heuristic methods to guide codon selection more efficiently. Together, these techniques could mitigate both algorithmic and hardware bottlenecks, extending the applicability of our hybrid framework to therapeutically sized mRNA sequences.

\bibliographystyle{myIEEEtran}
\bibliography{ref}

\begin{thebibliography}{10}
\providecommand{\url}[1]{#1}
\csname url@samestyle\endcsname
\providecommand{\newblock}{\relax}
\providecommand{\bibinfo}[2]{#2}
\providecommand{\BIBentrySTDinterwordspacing}{\spaceskip=0pt\relax}
\providecommand{\BIBentryALTinterwordstretchfactor}{4}
\providecommand{\BIBentryALTinterwordspacing}{\spaceskip=\fontdimen2\font plus
\BIBentryALTinterwordstretchfactor\fontdimen3\font minus \fontdimen4\font\relax}
\providecommand{\BIBforeignlanguage}[2]{{%
\expandafter\ifx\csname l@#1\endcsname\relax
\typeout{** WARNING: IEEEtran.bst: No hyphenation pattern has been}%
\typeout{** loaded for the language `#1'. Using the pattern for}%
\typeout{** the default language instead.}%
\else
\language=\csname l@#1\endcsname
\fi
#2}}
\providecommand{\BIBdecl}{\relax}
\BIBdecl

\bibitem{metkar2024tailor}
M.~Metkar, C.~S. Pepin, and M.~J. Moore, ``Tailor made: the art of therapeutic m{RNA} design,'' \emph{Nature Reviews Drug Discovery}, vol.~23, no.~1, pp. 67--83, 2024.

\bibitem{beaudoin20105}
J.-D. Beaudoin and J.-P. Perreault, ``5′-utr g-quadruplex structures acting as translational repressors,'' \emph{Nucleic acids research}, vol.~38, no.~20, pp. 7022--7036, 2010.

\bibitem{mauger2019mrna}
D.~M. Mauger, B.~J. Cabral, V.~Presnyak, S.~V. Su, D.~W. Reid, B.~Goodman, K.~Link, N.~Khatwani, J.~Reynders, M.~J. Moore \emph{et~al.}, ``mrna structure regulates protein expression through changes in functional half-life,'' \emph{Proceedings of the National Academy of Sciences}, vol. 116, no.~48, pp. 24\,075--24\,083, 2019.

\bibitem{andronescu2008rna}
M.~Andronescu, V.~Bereg, H.~H. Hoos, and A.~Condon, ``Rna strand: the rna secondary structure and statistical analysis database,'' \emph{BMC bioinformatics}, vol.~9, pp. 1--10, 2008.

\bibitem{leppek2022combinatorial}
K.~Leppek, G.~W. Byeon, W.~Kladwang, H.~K. Wayment-Steele, C.~H. Kerr, A.~F. Xu, D.~S. Kim, V.~V. Topkar, C.~Choe, D.~Rothschild \emph{et~al.}, ``Combinatorial optimization of mrna structure, stability, and translation for rna-based therapeutics,'' \emph{Nature communications}, vol.~13, no.~1, p. 1536, 2022.

\bibitem{li2024codonbert}
S.~Li, S.~Moayedpour, R.~Li, M.~Bailey, S.~Riahi, L.~Kogler-Anele, M.~Miladi, J.~Miner, F.~Pertuy, D.~Zheng \emph{et~al.}, ``Codonbert large language model for mrna vaccines,'' \emph{Genome research}, vol.~34, no.~7, pp. 1027--1035, 2024.

\bibitem{vostrosablin2024mrnaid}
N.~Vostrosablin, S.~Lim, P.~Gopal, K.~Brazdilova, S.~Parajuli, X.~Wei, A.~Gromek, D.~Prihoda, M.~Spale, A.~Muzdalo \emph{et~al.}, ``mrnaid, an open-source platform for therapeutic mrna design and optimization strategies,'' \emph{NAR Genomics and Bioinformatics}, vol.~6, no.~1, p. lqae028, 2024.

\bibitem{wayment2021theoretical}
H.~K. Wayment-Steele, D.~S. Kim, C.~A. Choe, J.~J. Nicol, R.~Wellington-Oguri, A.~M. Watkins, R.~A. Parra~Sperberg, P.-S. Huang, E.~Participants, and R.~Das, ``Theoretical basis for stabilizing messenger rna through secondary structure design,'' \emph{Nucleic acids research}, vol.~49, no.~18, pp. 10\,604--10\,617, 2021.

\bibitem{zhang2023algorithm}
H.~Zhang, L.~Zhang, A.~Lin, C.~Xu, Z.~Li, K.~Liu, B.~Liu, X.~Ma, F.~Zhao, H.~Jiang \emph{et~al.}, ``Algorithm for optimized mrna design improves stability and immunogenicity,'' \emph{Nature}, vol. 621, no. 7978, pp. 396--403, 2023.

\bibitem{gu2024derna}
X.~Gu, Y.~Qi, and M.~El-Kebir, ``Derna enables pareto optimal rna design,'' \emph{Journal of Computational Biology}, vol.~31, no.~3, pp. 179--196, 2024.

\bibitem{ward2025mrna}
M.~Ward, M.~Richardson, and M.~Metkar, ``mrna folding algorithms for structure and codon optimization,'' \emph{arXiv preprint arXiv:2503.19273}, 2025.

\bibitem{zhang2024resource}
H.~Zhang, A.~Sarkar, and K.~Bertels, ``A resource-efficient variational quantum algorithm for mrna codon optimization,'' \emph{arXiv preprint arXiv:2404.14858}, 2024.

\bibitem{chung2024quantum}
Y.~K. Chung, D.~Lee, J.~Lee, J.~Kim, D.~K. Park, and J.~Huh, ``Quantum-classical hybrid approach for codon optimization and its practical applications,'' \emph{bioRxiv}, pp. 2024--06, 2024.

\bibitem{sharp1987codon}
P.~M. Sharp and W.-H. Li, ``The codon adaptation index-a measure of directional synonymous codon usage bias, and its potential applications,'' \emph{Nucleic acids research}, vol.~15, no.~3, pp. 1281--1295, 1987.

\bibitem{tinoco1971estimation}
I.~Tinoco, O.~C. Uhlenbeck, and M.~D. Levine, ``Estimation of secondary structure in ribonucleic acids,'' \emph{Nature}, vol. 230, no. 5293, pp. 362--367, 1971.

\bibitem{fox_codon}
W.~R. Fox~DM, Branson~KM, ``{mRNA} codon optimization with quantum computers,'' \emph{PLoS One}, vol.~16, 2021. doi: 10.1371/journal.pone.0259101

\bibitem{plotkin2011synonymous}
J.~B. Plotkin and G.~Kudla, ``Synonymous but not the same: the causes and consequences of codon bias,'' \emph{Nature Reviews Genetics}, vol.~12, no.~1, pp. 32--42, 2011.

\bibitem{codtab}
G.~T. Nakamura~Y. and I.~T., ``Codon usage tabulated from the international dna sequence databases: status for the year 2000,'' \emph{Nucl. Acids Res.}, vol. 28(1):292, 2000.

\bibitem{alevras2024mrna}
D.~Alevras, M.~Metkar, T.~Yamamoto, V.~Kumar, T.~Friedhoff, J.-E. Park, M.~Takeori, M.~LaDue, W.~Davis, and A.~Galda, ``mrna secondary structure prediction using utility-scale quantum computers,'' in \emph{2024 IEEE International Conference on Quantum Computing and Engineering (QCE)}, vol.~1.\hskip 1em plus 0.5em minus 0.4em\relax IEEE, 2024, pp. 488--499.

\bibitem{gusfield2019integer}
D.~Gusfield, \emph{Integer Linear Programming in Computational and Systems Biology: An Entry-Level Text and Course}.\hskip 1em plus 0.5em minus 0.4em\relax Cambridge University Press, 2019.

\bibitem{ibmlicense}
{IBM ILOG CPLEX}. License information. Available from IBM: \url{https://www.ibm.com/products/ilog-cplex-optimization- studio}.

\bibitem{turner-mathews}
D.~H. Turner and D.~H. Mathews, ``{NNDB}: the nearest neighbor parameter database for predicting stability of nucleic acid secondary structure,'' \emph{Nucleic Acids Research}, vol.~38, no. {suppl\_1}, p. D280– D282, 2009.

\bibitem{nelder1965simplex}
J.~A. Nelder and R.~Mead, ``A simplex method for function minimization,'' \emph{The computer journal}, vol.~7, no.~4, pp. 308--313, 1965.

\bibitem{barkoutsos2020improving}
P.~K. Barkoutsos, G.~Nannicini, A.~Robert, I.~Tavernelli, and S.~Woerner, ``Improving variational quantum optimization using cvar,'' \emph{Quantum}, vol.~4, p. 256, 2020.

\bibitem{vienna}
R.~Lorenz, S.~H. Bernhart, C.~Höner~zu Siederdissen, H.~Tafer, C.~Flamm, P.~F. Stadler, and I.~L. Hofacker, ``Viennarna package 2.0,'' \emph{Algorithms for Molecular Biology}, vol. 6:1 26, 2011. doi: 10.1186/1748-7188-6-26

\end{thebibliography}
\end{document}